\documentclass[a4paper,11pt]{article}

\usepackage[T1]{fontenc}
\usepackage[utf8]{inputenc}
\usepackage[english]{babel}

\usepackage[a4paper,top=3cm,bottom=3cm,left=3cm,right=3cm]{geometry}

\usepackage{amsmath,amssymb,amsfonts,mathrsfs,stmaryrd}
\usepackage{braket}

\usepackage{array,booktabs,boldline,multirow,tabularx,colortbl,diagbox,makecell,fancybox}
\usepackage{xcolor}

\usepackage{graphicx}
\usepackage{pgf,tikz}
\usetikzlibrary{calc,positioning,shapes.geometric,shapes.symbols,shapes.misc,fit,shapes,arrows,arrows.meta,fadings,through}

\usepackage{listings}
\usepackage{fancyhdr}
\usepackage{setspace}
\usepackage[hidelinks]{hyperref}

\title{\textbf{Nonlocal Games Revisited: A Representation-Theoretic Path from Bell Locality to Quantum Pseudo-Telepathy}}

\usepackage[hidelinks]{hyperref}

\author{
Mustafa Mert \"Ozyılmaz\\
Sorbonne Universit\'e\\
\href{mailto:mustafa_mert.ozyilmaz@etu.sorbonne-universite.fr}{\nolinkurl{mustafa_mert.ozyilmaz@etu.sorbonne-universite.fr}}
\and
Ruchi Thareja\\
Sorbonne Universit\'e\\
\href{mailto:ruchi.thareja@etu.sorbonne-universite.fr}{\nolinkurl{ruchi.thareja@etu.sorbonne-universite.fr}}
\and
Houssam Nasser\\
Sorbonne Universit\'e\\
\href{mailto:houssam_eddine.nasser@etu.sorbonne-universite.fr}{\nolinkurl{houssam_eddine.nasser@etu.sorbonne-universite.fr}}
}

\date{April 10, 2026}

\begin{document}

\maketitle


\begin{abstract}
Nonlocal games provide a unified operational framework for studying the distinction between classical, quantum, and more general no-signaling correlations. In this work, we develop this perspective by connecting the Bell-locality framework to several complementary mathematical representations of nonlocal games and quantum strategies. We begin with local hidden-variable models, the CHSH inequality, and the role of Bell nonlocality as a device-independent witness of entanglement, and then introduce nonlocal games through the standard predicate/verifier formalism.

We next examine a set of representative examples, including XOR games, the GHZ game, graph-based coloring games, the Mermin--Peres magic square game, and Hardy's paradox as a related logical manifestation of nonlocality. Building on this foundation, we compare four closely related representation frameworks: conditional-probability and correlation descriptions, Bell-functional formulations, entangled-value optimization, and the quantum-operator approach together with the Navascu\'es--Pironio--Ac\'in (NPA) hierarchy. These viewpoints are then instantiated for the CHSH, magic square, and GHZ games, showing how each representation emphasizes a different aspect of the same underlying task.

Taken together, these examples show that nonlocal games can be studied simultaneously as operational tasks, geometric objects in correlation space, optimization problems over entangled resources, and operator-theoretic constructions. This multi-representation viewpoint clarifies the relation between Bell inequality violations, perfect quantum strategies, pseudo-telepathy, and semidefinite relaxations of quantum correlations.
\end{abstract}

\tableofcontents

\section{Overview of Bell Locality}

\subsection{Local Theories}

When performing an experiment such as generating pairs of spin-\(\frac{1}{2}\) particles and measuring their spins along different directions, the observed probabilities typically do not satisfy the factorization condition:

\[
P(a,b \mid x,y) \neq P(a \mid x)\,P(b \mid y).
\]

This indicates that the outcomes on both sides are not statistically independent. Even if the two systems are separated by a large distance potentially even in a space-like separated configuration, the existence of such correlations is not inherently paradoxical. In particular, it does not necessarily imply any direct causal influence of one measurement outcome on the other. Instead, these correlations could arise due to a common cause: a dependence established when the two systems interacted in the past.

A local theory formalizes this intuition by assuming that there exists a set of hidden variables, denoted by \(\lambda\), that encapsulate all relevant past influences. These variables account for the correlation between measurement outcomes, ensuring that once they are taken into account, the residual randomness in the outcomes is independent. Mathematically, this assumption of \textit{locality} is expressed as:

\[
P(a,b \mid x,y,\lambda) = P(a \mid x,\lambda)\,P(b \mid y,\lambda).
\]

This condition states that given full knowledge of \(\lambda\), the measurement outcomes \( a \) and \( b \) should be independent, meaning that the observed correlations arise purely from shared prior information rather than any direct influence between the measurements. \cite{Brunner2014}

\subsection{CHSH bound} 

Let us consider an experiment in which each observer, Alice and Bob, has only two measurement settings, denoted as \( x, y \in \{0,1\} \), and each measurement produces one of two possible outcomes, labeled as \( a, b \in \{-1, +1\} \). 

We define the expectation value of the product of their measurement outcomes for a given pair of measurement settings as:

\[
\langle a_x b_y \rangle = \sum_{a,b} P(a,b \mid x,y)\, a b\, 
\]

where the sum runs over all possible outcomes. Now, let us introduce the following linear combination of expectation values:

\[
S = \langle a_0 b_0 \rangle + \langle a_0 b_1 \rangle + \langle a_1 b_0 \rangle - \langle a_1 b_1 \rangle.
\]

This function depends on the joint probability distributions \( P(a,b \mid x,y) \). If these probabilities can be decomposed according to a \textit{local hidden variable model}, then we must satisfy the following inequality:

\[
\mid S \mid  \leq 2.
\]

This condition is known as the Clauser-Horne-Shimony-Holt (CHSH) inequality, introduced by Clauser et al. in 1969.

To derive this bound, we substitute the locality condition into the definition of the expectation values. Under the assumption of local realism, the expectation values can be expressed as an average over hidden variables \( \lambda \):

\[
\langle a_x b_y \rangle = \int d\lambda \, q(\lambda) \langle a_x \rangle_\lambda \langle b_y \rangle_\lambda,
\]

where \( q(\lambda) \) represents a probability distribution over hidden variables, and the local expectation values are given by:

\[
\langle a_x \rangle_\lambda = \sum_a P(a \mid x, \lambda)\,a\, \quad \langle b_y \rangle_\lambda = \sum_b P(b \mid y, \lambda)\,b\,.
\]

Both of these local expectation values must lie within the interval \([-1,1]\). Substituting this into our expression for \( S \), we obtain:

\[
S = \int d\lambda \, q(\lambda) S_\lambda,
\]

where the function \( S_\lambda \) is defined as:

\[
S_\lambda = \langle a_0 \rangle_\lambda \langle b_0 \rangle_\lambda + \langle a_0 \rangle_\lambda \langle b_1 \rangle_\lambda + \langle a_1 \rangle_\lambda \langle b_0 \rangle_\lambda - \langle a_1 \rangle_\lambda \langle b_1 \rangle_\lambda.
\]

Since \( \langle a_0 \rangle_\lambda \) and \( \langle a_1 \rangle_\lambda \) are constrained within \([-1,1]\), we analyze the worst-case scenario for \( S_\lambda \). By rewriting it as:

\[
 \mid S_\lambda \mid\leq |\langle b_0 \rangle_\lambda + \langle b_1 \rangle_\lambda| + |\langle b_0 \rangle_\lambda - \langle b_1 \rangle_\lambda|,
\]

we can assume, without loss of generality, that \( \langle b_0 \rangle_\lambda \geq \langle b_1 \rangle_\lambda \geq 0 \), leading to:

\[
\mid S_\lambda  \mid = 2 \langle b_0 \rangle_\lambda \leq 2.
\]

Since \( S \) is obtained as an average over \( S_\lambda \), it follows that:

\[
\mid S \mid = \int d\lambda \, q(\lambda) S_\lambda \leq 2.
\]

This confirms that any theory based on local hidden variables must satisfy the CHSH inequality, establishing a strict upper bound on the correlations that can arise in a local realistic framework. \cite{Brunner2014,CHSH1969}

\subsection{Nonlocality as a method of Entanglement Detection}

The unique capabilities of entangled quantum systems, which diverge significantly from classical systems, have been acknowledged for over a century. While classical physics provides a robust framework for many phenomena, it falls short in explaining certain quantum behaviors. Notably, when presented solely with the statistical outcomes from both a genuine quantum experiment and a classical simulation, distinguishing between the two based on raw measurement data alone is challenging. For instance, the data produced by a Stern-Gerlach experiment can be replicated through classical simulations; the inherent "non-classicality" becomes apparent only when considering that a magnetic moment is being measured in relation to the magnetic field gradient's direction.

\indent

In contrast, Bell nonlocality offers a scenario where the distinction between quantum experiments and classical simulations can be made directly from measurement data, independent of the specific physical systems or measurement details involved. This characteristic is known as device independence. In this framework, the violation of Bell inequalities serves as a robust indicator of entanglement, resilient to experimental imperfections. As long as a violation is observed, it guarantees, irrespective of implementation specifics, that the systems in question are entangled. This aspect is particularly significant because entanglement underpins numerous protocols in quantum information science, especially in quantum cryptography. Consequently, device-independent assessments pave the way for evaluating the performance of these protocols without relying on detailed knowledge of the devices used. \cite{Brunner2014} \\

The Bell-inequality viewpoint provides a powerful way to detect nonclassical correlations directly from observed statistics. A closely related but more operational language is obtained by recasting the same physical tension as a task between a referee and separated players. This leads to the framework of nonlocal games, where the central question is no longer only whether a Bell inequality can be violated, but also how different physical resources affect the players' optimal success probability. \cite{Brunner2014,Cleve2004}

\section{Nonlocal Games}

\subsection{Predicate / Verifier Form of a Nonlocal Game}

A nonlocal game is most commonly defined through a \emph{verifier} (or referee) who interacts with multiple separated players. In the standard two-player setting, the players are called Alice and Bob. They are not allowed to communicate after receiving their questions, but they may agree on a strategy beforehand and may share randomness or entanglement depending on the model under consideration.

Formally, a two-player nonlocal game \(G\) is specified by the tuple
\[
G = (X,Y,A,B,\pi,V),
\]
where:
\begin{itemize}
    \item \(X\) and \(Y\) are the finite sets of questions for Alice and Bob;
    \item \(A\) and \(B\) are the finite sets of answers for Alice and Bob;
    \item \(\pi : X \times Y \to [0,1]\) is a probability distribution over question pairs \((x,y)\);
    \item \(V : A \times B \times X \times Y \to \{0,1\}\) is the \emph{verification predicate} (or winning predicate).
\end{itemize}

The game proceeds as follows:
\begin{enumerate}
    \item The verifier samples a pair of questions \((x,y)\) according to the distribution \(\pi(x,y)\).
    \item The verifier sends \(x\) to Alice and \(y\) to Bob.
    \item Alice replies with an answer \(a \in A\), and Bob replies with an answer \(b \in B\).
    \item The players win if
    \[
    V(a,b,x,y)=1,
    \]
    and lose otherwise.
\end{enumerate}

Thus, the predicate \(V\) completely specifies the winning condition of the game: for every possible question pair \((x,y)\) and answer pair \((a,b)\), it determines whether that transcript is accepted by the verifier.

\paragraph{Strategies:}
A strategy for the players specifies how they choose their answers upon receiving their respective questions. In the most general probabilistic description, a strategy is represented by a family of conditional probability distributions
\[
P(a,b \mid x,y),
\]
where \(P(a,b \mid x,y)\) is the probability that Alice and Bob output \(a\) and \(b\) when they receive questions \(x\) and \(y\), respectively. The nature of the allowed set of correlations \(P(a,b \mid x,y)\) depends on the physical model:
\begin{itemize}
    \item in the classical case, the correlations arise from shared randomness;
    \item in the quantum case, they arise from local measurements on a shared entangled state;
    \item in more general no-signaling models, they are only required to satisfy the no-signaling constraints.
\end{itemize}

\paragraph{Winning probability:}
Given a strategy \(P(a,b \mid x,y)\), the success probability of the players in the game \(G\) is
\[
\omega(G;P)
=
\sum_{x \in X}\sum_{y \in Y} \pi(x,y)
\sum_{a \in A}\sum_{b \in B}
V(a,b,x,y)\, P(a,b \mid x,y).
\]
This expression averages the verifier's acceptance condition over all question pairs and all possible answers produced by the strategy.

The value of the game in a given model is the maximum winning probability over all strategies allowed in that model. For example, the classical value is
\[
\omega_c(G)=\sup_{P \in \mathcal{C}} \omega(G;P),
\]
where \(\mathcal{C}\) denotes the set of classical correlations, while the quantum value is
\[
\omega_q(G)=\sup_{P \in \mathcal{Q}} \omega(G;P),
\]
where \(\mathcal{Q}\) denotes the set of quantum correlations.

\paragraph{Interpretation:}
The predicate/verifier form is the most direct and operational way to define a nonlocal game. It separates the description of the task from the description of the resources used by the players. The tuple \((X,Y,A,B,\pi,V)\) defines only the rules of the game, while the distinction between classical, quantum, and no-signaling theories enters through the class of admissible strategies \(P(a,b \mid x,y)\). \cite{Cleve2004}

This predicate-based formulation serves as the standard starting point for other equivalent descriptions of nonlocal games that we will discuss in the later sections, such as the correlation form, Bell functional form, and operator-algebraic formulations. \cite{Cleve2004} \\

With these definitions in place, it will be useful to examine several representative examples before turning to more abstract formulations. The examples below are chosen to illustrate different structural regimes: CHSH as the simplest bipartite binary game, GHZ as a multipartite perfect quantum strategy, graph-coloring games as a broader combinatorial family, and the magic square game as a paradigmatic pseudo-telepathy task. Hardy's paradox will then serve as a nearby logical example that is not, strictly speaking, a standard nonlocal game, but still clarifies the broader landscape of nonlocality.

\subsection{CHSH Game and XOR Games }

\subsubsection{XOR Games}

An XOR game is a two-player nonlocal game where two players, commonly called Alice and Bob, receive inputs and produce binary outputs. The goal is to maximize their probability of winning based on a predefined condition. \cite{Cleve2004,Brunner2014}

\paragraph{General Definition of an XOR Game:} 

Formally, an XOR game \( G \) is defined by:
\begin{itemize}
    \item A finite set of inputs \( X \) and \( Y \) for Alice and Bob, respectively.
    \item A probability distribution \( \pi(x,y) \) over input pairs \( (x,y) \).
    \item A Boolean function \( f: X \times Y \to \{0,1\} \) that specifies the winning condition.
\end{itemize}

Each player responds with an output \( a, b \in \{0,1\} \), without communicating after receiving their inputs. They win if their outputs satisfy the condition:

\[
a \oplus b = f(x,y),
\]

where \( \oplus \) denotes addition modulo 2.

\paragraph{Classical Winning Probability:}

In a classical setting, the players can only use deterministic or probabilistic strategies based on pre-shared classical information. The best classical strategy leads to a maximum winning probability of:

\[
\omega_c(G) = \max_{a(x),\,b(y)} \sum_{x,y} \pi(x,y) \delta_{a(x) \oplus b(y), f(x,y)},
\]

where \( a(x) \) and \( b(y) \) are deterministic response functions for Alice and Bob.

\paragraph{Quantum Strategy of XOR Games:}

Alice and Bob share a pre-established entangled state \( |\psi\rangle \in \mathcal{H}_A \otimes \mathcal{H}_B \).
- Upon receiving inputs \( x \) and \( y \), they perform local binary measurements represented by observables:
  \[
  A_x \text{ on } \mathcal{H}_A, \quad B_y \text{ on } \mathcal{H}_B,
  \]
  where \( A_x^2 = B_y^2 = \mathbb{I} \) and the eigenvalues of \( A_x \) and \( B_y \) lie in \( \{\pm1\} \).
 The outcomes \( a, b \in \{0,1\} \) are obtained from these binary measurement results.

The quantum winning probability is given by:

\[
\omega_q(G) = \sup_{\ket{\psi}, A_x, B_y} \sum_{x,y} \pi(x,y) \frac{1 + (-1)^{f(x,y)} \langle \psi | A_x \otimes B_y | \psi \rangle}{2}.
\]

\paragraph{Tsirelson's Bound and Quantum Advantage} 

Using Tsirelson’s theorem, the quantum value of XOR games can be characterized through an optimization over inner products, which can be formulated as a semidefinite program. Specifically, for the CHSH game (a fundamental XOR game that we will discuss in more detail later), the maximum quantum value is:

\[
\omega_q(G) = \frac{1}{2} + \frac{1}{2\sqrt{2}} \approx 0.854.
\]

This exceeds the best classical strategy, which achieves at most \( 0.75 \), demonstrating quantum nonlocality. \cite{Cleve2004,Brunner2014}

\subsection{GHZ Game (Greenberger-Horne-Zeilinger Game)}

The Greenberger–Horne–Zeilinger (GHZ) game is a paradigmatic example of a nonlocal game that reveals the incompatibility of classical local realism with quantum mechanics without requiring statistical inequalities. It involves three players:Alice, Bob, and Charlie, who cannot communicate during the game but may share entanglement beforehand. \cite{Brassard2005,Mermin1990}

\subsubsection*{Game Structure}

Each player receives a binary input bit $x_i \in \{0,1\}$ with the promise that:
\[
x_1 \oplus x_2 \oplus x_3 = 0,
\]
i.e., the sum of inputs is even. The valid input combinations are:
\[
(0,0,0),\quad (0,1,1),\quad (1,0,1),\quad (1,1,0).
\]

Each player must respond with an output bit $a_i \in \{0,1\}$. \\

We relate the binary outputs to measurement outcomes \(m_i \in \{+1,-1\}\) via \( m_i = (-1)^{a_i} \), where \(m_i\) denotes the eigenvalue obtained from the corresponding Pauli measurement.

\subsubsection*{Winning Condition}

The players win the game if:
\[
a_1 \oplus a_2 \oplus a_3 = x_1 \lor x_2 \lor x_3.
\]

This translates to:
\begin{itemize}
  \item If $(x_1, x_2, x_3) = (0,0,0)$, then the output parity must be even.
  \item Otherwise, the output parity must be odd.
\end{itemize}

\subsubsection*{Classical and Quantum Strategies}

Using any classical strategy (with or without shared randomness), the players can achieve at most a $75\%$ success rate:
\[
\omega_c(G) = \frac{3}{4}.
\]

In contrast, if the players share the entangled GHZ state:
\[
\ket{\mathrm{GHZ}} = \frac{1}{\sqrt{2}} (\ket{000} + \ket{111}),
\]
and each performs a measurement determined by their input bit ($X$ if $x_i = 0$, $Y$ if $x_i = 1$), they can win the game with certainty:
\[
\omega_q(G) = 1.
\]

\subsubsection*{Significance}

The GHZ game demonstrates quantum nonlocality in a deterministic way. It reveals a logical contradiction with local realism. It is a central tool in quantum foundations, quantum networks, and multipartite entanglement studies.

\subsubsection*{Operator Formalism}

The quantum win conditions align with expectation values of joint observables:
\begin{align*}
\langle X \otimes X \otimes X \rangle &= +1, \\
\langle X \otimes Y \otimes Y \rangle &= -1, \\
\langle Y \otimes X \otimes Y \rangle &= -1, \\
\langle Y \otimes Y \otimes X \rangle &= -1.
\end{align*}

These correlations arise from the stabilizer properties of the GHZ state, showcasing perfect quantum correlations that defy classical explanation. \cite{Brassard2005,Mermin1990}

\subsection{Nonlocal Coloring Games (Graph-Based Games)}

Given a graph $G = (V, E)$ and a positive integer $c$, the nonlocal coloring game is defined as follows: \\

     The referee samples a pair $(u,v)$ from a distribution $\pi(u,v)$ supported on pairs with either $u=v$ or $(u,v)\in E$. Alice receives $u$ and Bob receives $v$, and each outputs a color in $[c]=\{1,\dots,c\}$. They win if they output the same color when $u=v$, and different colors when $(u,v)\in E$. \\

\textbf{Winning condition:}
\begin{itemize}
    \item If $u = v$, then Alice and Bob must output the same color: $a = b$
    \item If $(u, v) \in E$, then they must output different colors: $a \neq b$
\end{itemize}

This game models the constraint satisfaction problem of $c$-coloring a graph, where adjacent vertices must receive different colors. The goal is to determine whether such a coloring can be simulated in a distributed, possibly quantum, setting. \cite{Cameron2007}

\subsubsection{Classical and Quantum Values}

Let:
\begin{itemize}
    \item $\omega_c(G, c)$ be the classical value: the maximum win probability using classical strategies
    \item $\omega_q(G, c)$ be the quantum value: the maximum win probability using shared entanglement

\end{itemize}

The chromatic number $\chi(G)$ is the smallest $c$ such that $\omega_c(G,c) = 1$. The quantum chromatic number $\chi_q(G)$ is the smallest $c$ such that $\omega_q(G,c) = 1$. We always have:
\[
\chi_q(G) \leq \chi(G)
\]

Quantum resources can reduce the coloring number in some cases.

\subsubsection{Classical Strategies and Chromatic Number}

The classical value $\omega_c(G,c) = 1$ if and only if the graph $G$ is classically $c$-colorable, i.e., there exists a map:
\[
f : V \rightarrow [c]
\]
such that $f(u) \neq f(v)$ for all $(u,v) \in E$. The chromatic number $\chi(G)$ is the minimum such $c$.

Provided that every relevant test pair $(u,u)$ and $(u,v)\in E$ occurs with nonzero probability, one has
\[
\omega_c(G,c)=1 \iff c\ge \chi(G).
\]

For $c < \chi(G)$, even with optimal classical strategies, players will necessarily make errors with nonzero probability.

\subsubsection{Quantum Strategies and Quantum Chromatic Number}

Suppose Alice and Bob share a quantum state $\rho$ and perform measurements $\{M^u_a\}$ and $\{N^v_b\}$ for each vertex $u, v \in V$, where:

\begin{itemize}
    \item $M^u_a$, $N^v_b$ are POVMs (positive operator-valued measures),
    \item $a, b \in [c]$ are measurement outcomes (colors).
\end{itemize}

Their winning probability is given by:
\[
P_{\text{win}} = \sum_{(u,v)} \pi(u,v) \sum_{a,b} \text{Tr}[(M^u_a \otimes N^v_b)\rho] \cdot V(a,b \mid u,v)
\]

If there exists such a strategy with $P_{\text{win}} = 1$, then $G$ is said to be quantum $c$-colorable. The quantum chromatic number $\chi_q(G)$ is the minimal $c$ for which such a perfect quantum strategy exists.

It is known that:
\[
\chi_q(G) \leq \chi(G)
\]
and strict inequality is possible that is, quantum resources can reduce the coloring number. \cite{Cameron2007}

\subsection{Magic Square Game}

The  Mermin-Peres Magic Square Game is a two-player nonlocal game that demonstrates quantum advantage. The game is particularly interesting because quantum strategies allow Alice and Bob to win with certainty, while classical strategies fail. \cite{Brassard2005,Peres1990,Mermin1990}

\subsubsection{The Structure of the Magic Square Game}

 \textbf{Inputs}: 
    \begin{itemize}
        \item Alice receives a row index \( x \in \{1,2,3\} \).
        \item Bob receives a column index \( y \in \{1,2,3\} \).
    \end{itemize}
     \textbf{Outputs}:
    \begin{itemize}
        \item Alice outputs three bits \( (a_1, a_2, a_3) \) corresponding to her row.
        \item Bob outputs three bits \( (b_1, b_2, b_3) \) corresponding to his column.
    \end{itemize}

\paragraph{Constraints:}
The following parity conditions must be satisfied:
\begin{itemize}
    \item Alice's row sum must be even:
    \[
    a_1 \oplus a_2 \oplus a_3 = 0.
    \]
    \item Bob's column sum must be odd:
    \[
    b_1 \oplus b_2 \oplus b_3 = 1.
    \]
\end{itemize}

\paragraph{Winning Condition:}
Alice and Bob win if their outputs match at the intersection of Alice’s row and Bob’s column:
\[
a_y = b_x.
\]

\subsubsection{Classical Strategies and Their Limitation}
In a classical deterministic strategy, Alice and Bob must preassign values to all nine positions in a \( 3 \times 3 \) grid. However, due to the parity constraints:
\begin{itemize}
    \item Alice’s rows must sum to even.
    \item Bob’s columns must sum to odd.
\end{itemize}
\textbf{No classical assignment can satisfy both conditions simultaneously.} Thus, classical strategies can win the game at most with probability:
\[
\omega_c(G) = \frac{8}{9} \approx 0.89.
\]
\cite{Brassard2005}

\subsubsection{Quantum Winning Strategy}

\paragraph{Shared Quantum State:}
A perfect quantum strategy uses a maximally entangled state of local dimension $4$, which can be realized as two shared EPR pairs. One convenient choice is
\[
\ket{\psi} = \ket{\Phi^+}^{\otimes 2},
\qquad
\ket{\Phi^+} = \frac{1}{\sqrt{2}}(\ket{00}+\ket{11}).
\]

\paragraph{Quantum Measurements:}
Alice and Bob measure commuting two-qubit Pauli observables arranged in the standard magic-square table. A common choice is
\[
\begin{array}{ccc}
Z\otimes \mathbb{I} & \mathbb{I}\otimes Z & Z\otimes Z \\
\mathbb{I}\otimes X & X\otimes \mathbb{I} & X\otimes X \\
-Z\otimes X &- X\otimes Z & Y\otimes Y
\end{array}
\]
with the convention that the product of observables in every row is $+\mathbb{I}$, while the product in the first two columns is $+\mathbb{I}$ and in the last column is $-\mathbb{I}$ up to equivalent sign conventions. Alice measures the row specified by her input, Bob measures the column specified by his input, and they output the corresponding eigenvalues converted into bits.

Because the row and column observables commute within each question and agree on the intersection entry, the players can satisfy the parity constraints and the consistency condition simultaneously. Hence,
\[
\omega_q(G)=1.
\]

\subsubsection{Why the Classical Strategy Fails}

The impossibility of a perfect classical strategy is an algebraic parity contradiction. If one preassigns deterministic values to all nine cells of the square, then the product of all row parities must equal the product of all column parities, because both equal the product of all nine entries. But the rules demand that all row parities multiply to $+1$ while the column parities multiply to $-1$. This contradiction proves that no classical assignment can satisfy all constraints simultaneously.

Therefore the magic square game gives a clean example of pseudo-telepathy: quantum players can win with certainty, whereas classical players cannot.

\paragraph{Contextuality and the Mermin--Peres Square:}
The magic square construction is closely related to quantum contextuality. Roughly speaking, contextuality means that the outcome assigned to an observable cannot be understood as a pre-existing value that is independent of which other compatible observables are measured alongside it. The Mermin--Peres square provides a state-independent proof of this phenomenon: the observables can satisfy the required product relations within each commuting row or column, but there is no single global noncontextual assignment of values to all entries that is consistent with all these relations simultaneously. In this sense, the magic square game is a nonlocal-game realization of the same contextuality obstruction. \cite{Peres1990,Mermin1990,Brassard2005}

\subsection{Related Concepts: Hardy's Paradox as a Logical Form of Nonlocality}

While Hardy's paradox is not a standard nonlocal game in the technical sense, it provides a game-like logical demonstration of quantum nonlocality without the use of Bell inequalities.

\begin{itemize}
    \item \textbf{Setup:} Two parties, Alice and Bob, are associated with two overlapping Mach--Zehnder interferometers, traditionally described using an electron and a positron.
    
    \item \textbf{Key logical structure:}
    \begin{itemize}
        \item If both particles are simultaneously in the overlapping arms, annihilation occurs.
        \item A click at Alice's dark detector implies that Bob's particle must have been in the overlapping arm.
        \item A click at Bob's dark detector implies that Alice's particle must have been in the overlapping arm.
    \end{itemize}
    
    \item \textbf{Paradoxical quantum outcome:} Quantum mechanics nevertheless predicts a nonzero probability that both dark detectors click in the same run, even though classical local reasoning would rule this out.
    
    \item \textbf{Quantum feature:} In the original interferometer version, this paradoxical joint event occurs with probability \(1/16\), while optimized Hardy-type constructions can raise this success probability to about \(9\%\).
\end{itemize}

Unlike Bell inequality tests, Hardy's paradox shows nonlocality through a logical contradiction between local realism and a quantum event that occurs with nonzero probability. For this reason, it is often described as ``nonlocality without inequalities.'' \cite{Hardy1993,Brunner2014}

The examples in the previous section highlight that nonlocal games can be introduced through concrete rules and winning conditions, but they can also be studied through several equivalent mathematical viewpoints. In particular, one may focus on the induced conditional probabilities, on linear Bell functionals, on optimization over entangled strategies, or on operator-algebraic descriptions of measurements. The next section organizes these viewpoints systematically before we return to the canonical games and try to rewrite each of them in these different forms.

\section{Different Forms of Nonlocal Game Representations}

\subsection{Correlation Matrix Form}

\label{sec:correlation_matrix}

In the context of nonlocal games, a correlation representation compactly describes the joint probabilities \( P(a,b \mid x,y) \) of players' outputs given their inputs. More precisely, \( P(a,b \mid x,y) \) represents the correlations generated by a particular strategy in a nonlocal game. This form is widely used in quantum information theory, particularly for studying Bell inequalities. \cite{Brunner2014,Cleve2004}

\subsubsection{Structure of the Correlation Matrix}
For a two-player binary-input binary-output game (inputs \( x,y \in \{0,1\} \), outputs \( a,b \in \{0,1\} \)), the conditional probabilities may be arranged as:

\[
\mathbf{P} = \begin{pmatrix}
P(0,0 \mid 0,0) & P(0,1 \mid 0,0) & P(1,0 \mid 0,0) & P(1,1 \mid 0,0) \\
P(0,0 \mid 0,1) & P(0,1 \mid 0,1) & P(1,0 \mid 0,1) & P(1,1 \mid 0,1) \\
P(0,0 \mid 1,0) & P(0,1 \mid 1,0) & P(1,0 \mid 1,0) & P(1,1 \mid 1,0) \\
P(0,0 \mid 1,1) & P(0,1 \mid 1,1) & P(1,0 \mid 1,1) & P(1,1 \mid 1,1) \\
\end{pmatrix}
\]

For binary-output scenarios, one often also defines the correlators
\[
E_{xy} = P(a=b \mid x,y) - P(a \neq b \mid x,y),
\]
which can be arranged into the matrix
\[
\mathbf{E} = \begin{pmatrix}
E_{00} & E_{01} \\
E_{10} & E_{11}
\end{pmatrix}.
\]

However, this correlator matrix does not in general fully specify the full probability distribution \( P(a,b \mid x,y) \); in general, the local marginals must also be known.

\subsubsection{Significance}
\begin{itemize}
\item The full distribution \( P(a,b \mid x,y) \) distinguishes classical, quantum, and no-signaling correlations.
\item The correlators \( E_{xy} \) are especially useful in binary-output Bell scenarios such as CHSH and other XOR games.
\item These representations are fundamental tools for analyzing Bell nonlocality.
\item They are also used in device-independent protocols (e.g., randomness certification). \cite{Brunner2014}
\end{itemize}

\subsection{Bell Functionals and Dual Characterization of Nonlocality}

\label{sec:bell_functional}

The Bell functional form provides a linear framework to derive inequalities that distinguish classical, quantum, and no-signaling correlations. It is central to studying nonlocality in device-independent scenarios. \cite{Brunner2014}

\subsubsection{Mathematical Formulation}
For a bipartite scenario with inputs \(x \in \mathcal{X}\), \(y \in \mathcal{Y}\) and outputs \(a \in \mathcal{A}\), \(b \in \mathcal{B}\), the general Bell functional is:

\[
\mathcal{B} = \sum_{a,b,x,y} \alpha_{a,b,x,y} \cdot P(a,b \mid x,y)
\]

where \(\alpha_{a,b,x,y}\) are real coefficients. A Bell inequality is then obtained by comparing this quantity to a bound satisfied by all local correlations, for example
\[
\mathcal{B} \leq \beta_{\mathrm{L}}.
\]

In binary-output scenarios, some Bell functionals can also be written using correlators \(E_{xy} = P(a=b \mid x,y) - P(a \neq b \mid x,y)\):

\[
\mathcal{B} = \sum_{x,y} \beta_{xy} E_{xy}
\]

This correlator form is not fully general, but applies to important cases such as CHSH and other XOR-type games.

\subsubsection{Dual Formulation}
The dual perspective frames Bell inequalities as linear constraints on the local set of correlations:

\begin{itemize}
\item \textbf{Primal}: Determine whether \(P(a,b \mid x,y)\) admits a decomposition into local deterministic strategies
\item \textbf{Dual}: Find coefficients \(\alpha_{a,b,x,y}\) defining a Bell inequality that separates a nonlocal behavior from the local polytope
\end{itemize}

\subsubsection{Significance}
\begin{itemize}
\item Tests fundamental limits of local realism
\item Certifies quantum advantage in device-independent protocols
\item Provides a linear programming approach to characterize the local set of correlations \cite{Brunner2014}
\end{itemize}

\begin{table}[h]
\centering
\caption{CHSH Bell Functional Comparison}
\begin{tabular}{lll}
\toprule
\textbf{Theory} & \textbf{Bound} & \textbf{Interpretation} \\
\midrule
Classical (LHV) & \(\leq 2\) & Local hidden variables \\
Quantum & \(\leq 2\sqrt{2}\) & Entanglement advantage \\
No-signaling & \(\leq 4\) & Relativistic causality \\
\bottomrule
\end{tabular}
\end{table}

\subsection{Entangled (Quantum) Value Form}

\label{sec:extended_value}

The entangled value form quantifies the advantage of quantum strategies over classical ones by optimizing over shared entangled states and local measurements. \cite{Cleve2004}

\subsubsection{Classical vs. Quantum Values}

For a nonlocal game $G$ with:
\begin{itemize}
\item Inputs: $x \in \mathcal{X}$, $y \in \mathcal{Y}$
\item Outputs: $a \in \mathcal{A}$, $b \in \mathcal{B}$
\item Winning condition: $V(a,b \mid x,y) \in \{0,1\}$
\end{itemize}

Classical value:

\[
\omega_c(G)=\max_{P\in\mathcal{L}} \sum_{x,y}\pi(x,y)\sum_{a,b} V(a,b \mid x,y)\,P(a,b \mid x,y).
\]

The entangled value is
\[
\omega_q(G)=\sup_{\ket{\psi},\{A_a^x\},\{B_b^y\}}
\sum_{x,y}\pi(x,y)\sum_{a,b}
V(a,b \mid x,y)\,\bra{\psi}A_a^x\otimes B_b^y\ket{\psi},
\]
where for each input $x$, $\{A_a^x\}_a$ is a POVM on Alice's space, and for each input $y$, $\{B_b^y\}_b$ is a POVM on Bob's space.

\subsubsection{Operator-Theoretic Representation}

For fixed measurements, define the game operator
\[
\mathcal{G}_G(\{A_a^x\},\{B_b^y\})
=
\sum_{x,y}\pi(x,y)\sum_{a,b}
V(a,b \mid x,y)\,A_a^x\otimes B_b^y.
\]

Then
\[
\omega_q(G)
=
\sup_{\{A_a^x\},\{B_b^y\}}
\lambda_{\max}\!\bigl(\mathcal{G}_G(\{A_a^x\},\{B_b^y\})\bigr).
\]

Since $V(a,b \mid x,y)\ge 0$ and $\pi(x,y)\ge 0$, the operator above is positive semidefinite, so equivalently
\[
\omega_q(G)
=
\sup_{\{A_a^x\},\{B_b^y\}}
\left\|
\mathcal{G}_G(\{A_a^x\},\{B_b^y\})
\right\|.
\]

\subsection{Quantum Operator Form and the NPA Hierarchy}

The quantum-operator formulation describes a nonlocal game in terms of a shared quantum state and local measurement operators. It provides the standard mathematical framework for defining quantum strategies and for expressing the optimal quantum winning probability as an optimization problem over states and measurements. The NPA hierarchy then gives a systematic semidefinite-programming relaxation of this optimization, yielding computable upper bounds on the achievable game value. \cite{Cleve2004,NPA2008}

\subsubsection{Quantum Strategies for a Nonlocal Game}

Consider a bipartite nonlocal game
\[
G=(X,Y,A,B,\pi,V),
\]
where \(X,Y\) are the question sets, \(A,B\) are the answer sets, \(\pi(x,y)\) is the input distribution, and \(V(a,b\mid x,y)\) is the winning predicate.

A quantum strategy for this game is specified by:
\begin{itemize}
    \item a shared bipartite state \(\rho\) on \(\mathcal{H}_A \otimes \mathcal{H}_B\),
    \item a family of measurement operators \(\{A_a^x\}_{a \in A}\) for each question \(x \in X\),
    \item a family of measurement operators \(\{B_b^y\}_{b \in B}\) for each question \(y \in Y\),
\end{itemize}
where each family forms a POVM:
\[
A_a^x \succeq 0, \qquad \sum_{a \in A} A_a^x = \mathbb{I},
\]
\[
B_b^y \succeq 0, \qquad \sum_{b \in B} B_b^y = \mathbb{I}.
\]

The corresponding conditional probabilities are
\[
P(a,b\mid x,y)=\mathrm{Tr}\!\bigl(\rho\,(A_a^x \otimes B_b^y)\bigr).
\]

The winning probability of this strategy is therefore
\[
\omega(G;\rho,A,B)
=
\sum_{x \in X}\sum_{y \in Y}\pi(x,y)
\sum_{a \in A}\sum_{b \in B}
V(a,b\mid x,y)\,
\mathrm{Tr}\!\bigl(\rho\,(A_a^x \otimes B_b^y)\bigr).
\]

The \emph{entangled value} of the game is defined as
\[
\omega_q(G)
=
\sup_{\rho,\{A_a^x\},\{B_b^y\}}
\omega(G;\rho,A,B).
\]

Thus, in the quantum-operator picture, computing the value of a nonlocal game becomes an optimization problem over shared quantum states and local measurements.

\subsubsection{Commuting-Operator Formulation}

For the purpose of upper bounds, it is often useful to consider the more general commuting-operator model. In this formulation, Alice's and Bob's measurement operators act on a common Hilbert space \(\mathcal{H}\), and one writes
\[
P(a,b\mid x,y)=\bra{\psi} M_a^x N_b^y \ket{\psi},
\]
where \(\ket{\psi} \in \mathcal{H}\), the operators satisfy
\[
M_a^x \succeq 0, \qquad \sum_a M_a^x=\mathbb{I},
\]
\[
N_b^y \succeq 0, \qquad \sum_b N_b^y=\mathbb{I},
\]
and, crucially,
\[
[M_a^x,N_b^y]=0
\qquad
\text{for all }a,b,x,y.
\]

This leads to the commuting-operator value
\[
\omega^{qc}(G)
=
\sup_{\ket{\psi},\{M_a^x\},\{N_b^y\}}
\sum_{x,y}\pi(x,y)\sum_{a,b}
V(a,b\mid x,y)\,
\bra{\psi} M_a^x N_b^y \ket{\psi}.
\]

Since every tensor-product strategy gives rise to a commuting-operator strategy, one always has
\[
\omega_q(G)\leq \omega^{qc}(G).
\]

\subsubsection{The NPA Hierarchy}

The NPA hierarchy provides a sequence of semidefinite relaxations for the commuting-operator optimization above. The idea is to replace the operators and state by a finite matrix of moments that must satisfy the algebraic constraints obeyed by any genuine quantum strategy.

At a fixed level of the hierarchy, one chooses a finite set of operator words
\[
S=\{\mathbb{I},\, M_a^x,\, N_b^y,\, M_a^xN_b^y,\, \ldots\}
\]
up to some bounded length. One then defines the moment matrix
\[
\Gamma_{ij}
=
\bra{\psi} S_i^{\dagger} S_j \ket{\psi}.
\]

This matrix must satisfy:

\begin{itemize}
    \item \textbf{Positive semidefiniteness:}
    \[
    \Gamma \succeq 0;
    \]

    \item \textbf{Normalization relations:}
    \[
    \sum_a M_a^x=\mathbb{I},
    \qquad
    \sum_b N_b^y=\mathbb{I};
    \]

    \item \textbf{Orthogonality relations} for projective measurements, when such a formulation is used:
    \[
    M_a^x M_{a'}^x=0 \quad (a\neq a'),
    \qquad
    N_b^y N_{b'}^y=0 \quad (b\neq b');
    \]

    \item \textbf{Commutation relations:}
    \[
    [M_a^x,N_b^y]=0.
    \]
\end{itemize}

In addition, certain entries of \(\Gamma\) correspond directly to observable probabilities, so the game objective becomes a linear function of moment-matrix entries. Each truncation level therefore defines a semidefinite program whose optimum gives an upper bound on \(\omega^{qc}(G)\), and hence also on \(\omega_q(G)\).

\subsubsection{Interpretation for Nonlocal Games}

The importance of the NPA hierarchy in the theory of nonlocal games is that it converts the problem
of bounding the best quantum winning probability into a tractable sequence of SDPs. Higher levels of the hierarchy impose more constraints and therefore give tighter bounds. In the limit, the hierarchy converges to the commuting-operator value:
\[
\lim_{k\to\infty}\omega^{(k)}(G)=\omega^{qc}(G),
\]
where \(\omega^{(k)}(G)\) denotes the optimum at level \(k\) of the hierarchy.

Therefore, the quantum-operator formulation gives the conceptual description of a quantum strategy, while the NPA hierarchy provides a practical method for bounding nonlocal game values and studying the set of correlations compatible with such strategies. \cite{NPA2008} \\

Having introduced the general representation frameworks, we will now specialize them to concrete games in the next section.

\section{CHSH Game Representations}
\label{sec:chsh_representations}

The CHSH game can be described through several closely related mathematical formulations.The probability/correlation form describes the observed behavior, the Bell functional evaluates the strength of nonlocality, the operator form gives an explicit quantum realization, and the NPA hierarchy provides a semidefinite-programming characterization of quantumly achievable correlations. \cite{Brunner2014,Cleve2004,NPA2008}

\subsection{Correlation Form}
\label{subsec:chsh_correlation}

For inputs \(x,y \in \{0,1\}\) and outputs \(a,b \in \{0,1\}\), the behavior of the game is described by the conditional probabilities \(P(a,b \mid x,y)\). For binary outputs, it is convenient to pass to correlators
\[
E_{xy} \;=\; \sum_{a,b\in\{0,1\}} (-1)^{a+b} P(a,b \mid x,y).
\]
Equivalently, \(E_{xy} = P(a=b \mid x,y) - P(a\neq b \mid x,y)\).

For the standard optimal quantum strategy, the local marginals are uniform, so the full probability distribution is determined by the correlators through
\[
P(a,b \mid x,y)
=
\frac{1}{4}\Bigl(1 + (-1)^{a+b} E_{xy}\Bigr).
\]

\begin{itemize}
\item \textbf{Probability matrix} \(\mathbf{P}\):
\[
\mathbf{P} =
\begin{pmatrix}
P(0,0 \mid 0,0) & P(0,1 \mid 0,0) & P(1,0 \mid 0,0) & P(1,1 \mid 0,0) \\
P(0,0 \mid 0,1) & P(0,1 \mid 0,1) & P(1,0 \mid 0,1) & P(1,1 \mid 0,1) \\
P(0,0 \mid 1,0) & P(0,1 \mid 1,0) & P(1,0 \mid 1,0) & P(1,1 \mid 1,0) \\
P(0,0 \mid 1,1) & P(0,1 \mid 1,1) & P(1,0 \mid 1,1) & P(1,1 \mid 1,1)
\end{pmatrix}.
\]

\item \textbf{Quantum-optimal correlators}:
\[
\mathbf{E}_{\text{quantum}}
=
\begin{pmatrix}
\frac{1}{\sqrt{2}} & \frac{1}{\sqrt{2}} \\
\frac{1}{\sqrt{2}} & -\frac{1}{\sqrt{2}}
\end{pmatrix}.
\]

\item \textbf{Quantum-optimal probabilities}: using the relation above, one obtains
\[
\mathbf{P}_{\text{quantum}}
=
\frac{1}{8}
\begin{pmatrix}
2+\sqrt{2} & 2-\sqrt{2} & 2-\sqrt{2} & 2+\sqrt{2} \\
2+\sqrt{2} & 2-\sqrt{2} & 2-\sqrt{2} & 2+\sqrt{2} \\
2+\sqrt{2} & 2-\sqrt{2} & 2-\sqrt{2} & 2+\sqrt{2} \\
2-\sqrt{2} & 2+\sqrt{2} & 2+\sqrt{2} & 2-\sqrt{2}
\end{pmatrix}.
\]
The first three rows correspond to positive correlator \(E_{xy}=1/\sqrt{2}\), so the even-parity outputs \((0,0)\) and \((1,1)\) are favored. The last row corresponds to \(E_{11}=-1/\sqrt{2}\), so the odd-parity outputs \((0,1)\) and \((1,0)\) are favored. \cite{Brunner2014}
\end{itemize}

\subsection{Bell Functional Form}
\label{subsec:chsh_bell}

In correlator language, the CHSH expression is the Bell functional
\[
S_{\mathrm{CHSH}}
=
E_{00}+E_{01}+E_{10}-E_{11}
=
\sum_{x,y\in\{0,1\}} \beta_{xy} E_{xy},
\]
with coefficient matrix
\[
\beta_{xy}
=
\begin{pmatrix}
1 & 1 \\
1 & -1
\end{pmatrix}.
\]

This functional separates different correlation sets:
\begin{align*}
\text{Classical: } & S_{\mathrm{CHSH}} \le 2, \\
\text{Quantum: } & S_{\mathrm{CHSH}} \le 2\sqrt{2}, \\
\text{No-signaling: } & S_{\mathrm{CHSH}} \le 4.
\end{align*}

Since the CHSH game has uniformly distributed inputs, the winning probability is directly related to the Bell value by
\[
\omega(\mathrm{CHSH})
=
\frac{1}{2} + \frac{1}{8} S_{\mathrm{CHSH}}.
\]
Thus the optimal quantum winning probability is obtained from the optimal Bell value \(S_{\mathrm{CHSH}}=2\sqrt{2}\). \cite{Brunner2014,CHSH1969}

\subsection{Quantum Operator Form and NPA Hierarchy}
\label{subsec:chsh_npa}

In the quantum formulation, a strategy consists of a bipartite state \(\ket{\psi}\) together with local observables \(A_0,A_1\) for Alice and \(B_0,B_1\) for Bob, each having eigenvalues \(\pm 1\). The correlators are then
\[
E_{xy}
=
\langle \psi \mid A_x \otimes B_y \mid \psi \rangle.
\]

A standard optimal choice is
\begin{align*}
A_0 &= \sigma_z, \qquad A_1 = \sigma_x, \\
B_0 &= \frac{\sigma_z + \sigma_x}{\sqrt{2}}, \qquad
B_1 = \frac{\sigma_z - \sigma_x}{\sqrt{2}},
\end{align*}
acting on the Bell state
\[
\ket{\psi}
=
\frac{\ket{00}+\ket{11}}{\sqrt{2}}.
\]
This realizes
\[
E_{00}=E_{01}=E_{10}=\frac{1}{\sqrt{2}},
\qquad
E_{11}=-\frac{1}{\sqrt{2}},
\]
and therefore
\[
S_{\mathrm{CHSH}} = 2\sqrt{2}.
\]

The associated Bell operator is
\[
\mathcal{B}_{\mathrm{CHSH}}
=
A_0 \otimes B_0
+
A_0 \otimes B_1
+
A_1 \otimes B_0
-
A_1 \otimes B_1,
\]
so that
\[
S_{\mathrm{CHSH}}
=
\langle \psi \mid \mathcal{B}_{\mathrm{CHSH}} \mid \psi \rangle.
\]

To characterize quantumly achievable correlations more systematically, one can use the \textbf{NPA hierarchy}. At the first level, one considers the monomial set
\[
\mathcal{S} = \{\mathbb{I}, A_0, A_1, B_0, B_1\}
\]
and builds the moment matrix
\[
\Gamma_{ij}
=
\langle \psi \mid s_i^\dagger s_j \mid \psi \rangle,
\qquad s_i,s_j \in \mathcal{S}.
\]
With the ordering \((\mathbb{I},A_0,A_1,B_0,B_1)\), this gives
\[
\Gamma
=
\begin{pmatrix}
1 & \langle A_0\rangle & \langle A_1\rangle & \langle B_0\rangle & \langle B_1\rangle \\
\langle A_0\rangle & 1 & \langle A_0A_1\rangle & \langle A_0B_0\rangle & \langle A_0B_1\rangle \\
\langle A_1\rangle & \langle A_1A_0\rangle & 1 & \langle A_1B_0\rangle & \langle A_1B_1\rangle \\
\langle B_0\rangle & \langle B_0A_0\rangle & \langle B_0A_1\rangle & 1 & \langle B_0B_1\rangle \\
\langle B_1\rangle & \langle B_1A_0\rangle & \langle B_1A_1\rangle & \langle B_1B_0\rangle & 1
\end{pmatrix}.
\]
The constraint \(\Gamma \succeq 0\), together with the operator relations \(A_x^2 = B_y^2 = \mathbb{I}\) and \([A_x,B_y]=0\), yields a semidefinite relaxation of the quantum set. For CHSH, this first nontrivial level already recovers the Tsirelson bound \(2\sqrt{2}\).

It is important to note that entries such as \(\langle A_0A_1\rangle\) or \(\langle B_0B_1\rangle\) are not directly observed correlators; they are auxiliary moment variables introduced by the relaxation. \cite{NPA2008}

\subsection{Extended Value (Entangled Value) Form}
\label{subsec:chsh_extended}

The entangled value of the CHSH game is the maximum winning probability over all shared entangled states and local measurements. In Bell-operator language, one may first define the optimal CHSH Bell value \cite{Cleve2004,Brunner2014}
\[
\beta_q(\mathrm{CHSH})
=
\sup_{\ket{\psi},\,A_x,\,B_y}
\langle \psi \mid \mathcal{B}_{\mathrm{CHSH}} \mid \psi \rangle
=
2\sqrt{2}.
\]
The corresponding game value is then
\[
\omega_q(\mathrm{CHSH})
=
\frac{1}{2} + \frac{1}{8}\,\beta_q(\mathrm{CHSH})
=
\frac{2+\sqrt{2}}{4}
\approx 0.8536.
\]

Thus, the Bell value \(2\sqrt{2}\) and the winning probability \((2+\sqrt{2})/4\) are different but directly related quantities. The former measures the maximal violation of the CHSH inequality, while the latter is the actual optimal success probability in the game. \cite{Cleve2004,Brunner2014}

The CHSH game is the cleanest setting in which the relationship between correlations, Bell inequalities, operator realizations, and game values can be seen explicitly. However, it remains a binary-output bipartite game, so some of its mathematical simplifications are special to that setting. To see how the same general framework extends beyond XOR-type scenarios, we next consider the magic square game, where the outputs are structured bit strings and the quantum advantage takes the stronger form of pseudo-telepathy.

\section{Magic Square Game Representations}
\label{sec:magic_square_representations}

\subsection{Conditional-Probability Form}
\label{subsec:magic_probability}

The magic square game is naturally described by the conditional probabilities
\[
P(a,b\mid x,y),
\]
where \(x \in \{1,2,3\}\) is Alice's row input, \(y \in \{1,2,3\}\) is Bob's column input, and the outputs are parity-constrained bit strings.

\begin{itemize}
\item \textbf{Inputs}: Alice receives a row index \(x \in \{1,2,3\}\) and Bob receives a column index \(y \in \{1,2,3\}\).

\item \textbf{Outputs}: Alice outputs a bit string \(a=(a_1,a_2,a_3)\in\{0,1\}^3\) satisfying
\[
a_1\oplus a_2\oplus a_3=0,
\]
and Bob outputs a bit string \(b=(b_1,b_2,b_3)\in\{0,1\}^3\) satisfying
\[
b_1\oplus b_2\oplus b_3=1.
\]

\item \textbf{Consistency condition}: the intersection bits must agree,
\[
a_y=b_x.
\]
\end{itemize}

For an ideal quantum strategy, all probability weight is concentrated on winning outputs, so for every input pair \((x,y)\), \cite{Brassard2005}
\[
\sum_{a,b} V(a,b\mid x,y)\,P(a,b\mid x,y)=1.
\]

\subsection{Bell Functional Form}
\label{subsec:magic_bell}

A Bell-functional presentation of the game value is
\[
\omega(G)=\sum_{x=1}^{3}\sum_{y=1}^{3}\pi(x,y)\sum_{a,b} V(a,b\mid x,y)\,P(a,b\mid x,y),
\]
where \(\pi(x,y)=1/9\) for the uniform version of the game and \(V(a,b\mid x,y)=1\) exactly when Alice's row parity is even, Bob's column parity is odd, and the consistency condition \(a_y=b_x\) holds.

For the standard magic square game,
\[
\omega_c(G)=\frac{8}{9},
\qquad
\omega_q(G)=1.
\]
\cite{Brassard2005}

\subsection{Quantum Operator Form}
\label{subsec:magic_operator}

A standard perfect quantum strategy uses local two-qubit Pauli observables arranged in the Mermin--Peres square
\[
\begin{array}{ccc}
Z\otimes \mathbb{I} & \mathbb{I}\otimes Z & Z\otimes Z \\
\mathbb{I}\otimes X & X\otimes \mathbb{I} & X\otimes X \\
-\,Z\otimes X & -\,X\otimes Z & Y\otimes Y
\end{array}
\]
with the property that observables in each row commute and observables in each column commute.

Alice's input \(x\) specifies which row is measured, and Bob's input \(y\) specifies which column is measured. Under the identification
\[
0 \leftrightarrow +1,
\qquad
1 \leftrightarrow -1,
\]
the row products are \(+\mathbb{I}\), giving Alice even parity, while the column products are \(-\mathbb{I}\), giving Bob odd parity. The common intersection observable ensures the consistency condition. \cite{Brassard2005,Peres1990,Mermin1990}

\subsection{Extended Value (Entangled Value) Form}
\label{subsec:magic_extended}

The entangled value is
\[
\omega_q(G)=\sup_{\ket{\psi},\{\Pi_a^x\},\{\Pi_b^y\}}
\sum_{x,y}\pi(x,y)\sum_{a,b}V(a,b\mid x,y)\,
\bra{\psi}\Pi_a^x\otimes \Pi_b^y\ket{\psi},
\]
where \(\{\Pi_a^x\}\) and \(\{\Pi_b^y\}\) are Alice's and Bob's measurement operators.

For the magic square game,
\[
\omega_q(G)=1.
\]
One optimal realization uses two shared EPR pairs,
\[
\ket{\psi}=\ket{\Phi^+}^{\otimes 2},
\]
together with the commuting Pauli measurements above. \cite{Brassard2005} \\

The magic square game shows that the representation framework remains meaningful even when one leaves the binary-output setting and moves to a task with richer local answer structure. The next example changes a different aspect of the problem: instead of remaining bipartite with more complicated outputs, we move to a multipartite game. The GHZ game is therefore a natural complement to both CHSH and the magic square game, since it highlights how perfect quantum strategies appear in the genuinely three-party setting.

\section{GHZ Game Representations}
\label{sec:ghz_representations}

\subsection{Support-Table Form}
\label{subsec:ghz_support}

Because the game has three players, the natural probability object is the tripartite conditional distribution
\[
P(a_1,a_2,a_3 \mid x_1,x_2,x_3),
\]
or equivalently the GHZ/Mermin support table.

\begin{itemize}
\item \textbf{Inputs}: each player receives $x_i \in \{0,1\}$ with the promise
\[
x_1 \oplus x_2 \oplus x_3 = 0.
\]
Thus the valid questions are $000,011,101,110$.

\item \textbf{Outputs}: each player returns $a_i \in \{0,1\}$.

\item \textbf{Winning condition}:
\[
a_1 \oplus a_2 \oplus a_3 =
\begin{cases}
0 & \text{for } 000, \\
1 & \text{for } 011,101,110.
\end{cases}
\]
\end{itemize}

For the ideal GHZ strategy, each valid question has exactly four winning output strings, each occurring with probability $1/4$, while all losing outputs occur with probability $0$. \cite{Brassard2005,Mermin1990}

\subsection{Bell Functional Form}
\label{subsec:ghz_bell}

The game value can be written as
\[
\omega(G)=\sum_{\vec{x}} \pi(\vec{x}) \sum_{\vec{a}} V(\vec{a},\vec{x}) P(\vec{a}\mid\vec{x}),
\]
with $\pi(\vec{x})=1/4$ on the four valid inputs.

After encoding the outputs as $\pm 1$ observables, the associated Bell functional is the Mermin expression
\[
M = X\otimes X\otimes X - X\otimes Y\otimes Y - Y\otimes X\otimes Y - Y\otimes Y\otimes X.
\]
It is related to the GHZ game value by
\[
\omega(G)=\frac{1}{2}+\frac{1}{8}\langle M\rangle.
\]

Local hidden-variable models satisfy
\[
\langle M \rangle \leq 2,
\]
whereas the GHZ state achieves
\[
\langle M \rangle = 4.
\]
Equivalently,
\[
\omega_c(G)=\frac{3}{4},
\qquad
\omega_q(G)=1,
\qquad
\omega_{ns}(G)=1.
\]
\cite{Brassard2005,Mermin1990}

\subsection{Quantum Operator Form}
\label{subsec:ghz_operator}

A perfect quantum strategy is obtained from the state
\[
\ket{\mathrm{GHZ}}=\frac{1}{\sqrt{2}}(\ket{000}+\ket{111})
\]
with local observables
\[
A_0=B_0=C_0=X,
\qquad
A_1=B_1=C_1=Y.
\]
The corresponding correlators are
\begin{align*}
\langle X \otimes X \otimes X \rangle &= +1, \\
\langle X \otimes Y \otimes Y \rangle &= -1, \\
\langle Y \otimes X \otimes Y \rangle &= -1, \\
\langle Y \otimes Y \otimes X \rangle &= -1.
\end{align*}
These are exactly the correlators entering the Mermin functional and they certify perfect play of the GHZ game. \cite{Brassard2005,Mermin1990}

\subsection{Extended Value (Entangled Value) Form}
\label{subsec:ghz_extended}

The entangled value is
\[
\omega_q(G)=
\sup_{\ket{\psi},\,\{\Pi_{a_1}^{x_1}\},\,\{\Pi_{a_2}^{x_2}\},\,\{\Pi_{a_3}^{x_3}\}}
\sum_{\vec{x}} \pi(\vec{x}) \sum_{\vec{a}} V(\vec{a},\vec{x})\,
\bra{\psi}\,
\Pi_{a_1}^{x_1}\otimes \Pi_{a_2}^{x_2}\otimes \Pi_{a_3}^{x_3}
\,\ket{\psi},
\]
where each $\{\Pi_{a_i}^{x_i}\}_{a_i}$ is a projective measurement (or, more generally, a POVM).

For the GHZ game,
\[
\omega_q(G)=1,
\]
achieved by the state
\[
\ket{\mathrm{GHZ}}=\frac{1}{\sqrt{2}}(\ket{000}+\ket{111}),
\]
with $x_i=0$ corresponding to measurement in the $X$ basis and $x_i=1$ corresponding to measurement in the $Y$ basis.

\section{Conclusion}

In this work, we moved from the Bell-locality viewpoint to the mathematical framework of nonlocal games, and then compared several mathematical ways of representing quantum and classical strategies. The central lesson is that these formulations are not separate theories, but different perspectives on the same underlying phenomena. The correlation viewpoint emphasizes observable behavior, the Bell-functional viewpoint emphasizes witnesses of nonlocality, the entangled-value viewpoint emphasizes optimization over physical resources, and the operator viewpoint emphasizes the measurement structure underlying quantum strategies.

The case studies of the CHSH, magic square, and GHZ games show how these representations adapt to different scenarios. CHSH provides the simplest bipartite binary example, the magic square game illustrates pseudo-telepathy and contextuality-related structure, and the GHZ game highlights deterministic multipartite nonlocality. Taken together, these examples show both the unity of the nonlocal-game framework and the diversity of the quantum phenomena it captures.

\section*{Acknowledgements}

This work was initially carried out as a project for the Quantum Information Master's program at Sorbonne Universit\'e. The author sincerely thanks Prof.\ Marco T\'ulio Quintino for his guidance and support.

\bibliographystyle{plain} 
\bibliography{references}

\end{document}